\documentclass[doublecol]{epl2}

\title{Localized states and interaction induced delocalization in Bose gases with quenched disorder}
\shorttitle{Title} 

\author{G.M. Falco\inst{1} \and T. Nattermann\inst{1} \and V.L. Pokrovsky\inst{2,3}}
\shortauthor{F. Author \etal}

\institute{
  \inst{1} Institut f\"{u}r Theoretische Physik, Universit\"{a}t zu K\"{o}ln,
Z\"{u}lpicher Str. 77 D-50937 K\"{o}ln, Germany\\
  \inst{2} Dept. of Physics, Texas A\&M University, College Station, TX 77843-4242\\
  \inst{3} Landau Institute for Theoretical Physics, Chernogolovka, Moscow District
142432, Russia
}

\pacs{03.75.Hh}{03.75.Kk}

\abstract{
Very diluted Bose gas placed into a disordered environment falls into a fragmented
localized state. At some critical density the repulsion between particles overcomes the disorder. 
The gas transits into a coherent superfluid state.
In this article the geometrical and energetic characteristics of the localized
state at zero temperature and the critical density at which the quantum phase
transition from the localized to the superfluid state proceeds are found.}

\begin{document}

\maketitle

\section{Introduction}
The interplay between interaction and disorder is an important paradigm of condensed matter physics. In 1958 Anderson\cite{Anderson58} showed that in disordered solids a non-interacting electron may become localized due to the quantum interference. A phenomenological theory of localization\cite{gangof4,review} concluded that non-interacting electrons in one and two dimensions are always localized. In three dimensions the localized and extended states are separated by the mobility edge.  States with energy significantly below this edge in 3 dimensions are strongly localized. They appear
in rare fluctuations of the quenched random potential\cite{ILifshitz,zittartz,halperin-lax}.
These instanton-type states broaden and eventually overlap with growing energy.
A system of non-interacting fermions in the random potential transits from
the insulator to metal state when its Fermi energy exceeds the mobility edge.
Thus, the Pauli principle delocalizes fermions
in 3 dimensions, but leave them localized in lower dimensions. The common
belief is that the repulsive interaction suppresses the localization. So
far this problem was studied only in the limit of a weak disorder\cite{altshuler+aronov,
finkelstein}. Therefore,
the interaction induced delocalization transition remains beyond the frameworks
of the theory. The metal-insulator transition in 2 dimensions was observed in
experiments\cite{
kravchenko-b} suggesting the decisive role of interaction.
The physical picture changes drastically for bosons. The non-interacting bosons
condense at a single-particle state with the lowest energy. In a homogeneous system
it leads to a coherent quantum state known as the Bose-Einstein condensate (BEC). Examples are superfluid phases of He\cite{Leggett}, superconductors \cite{Thinkham},   BEC  of ultra-cold alkali atoms\cite{Ketterle,Dalfovo} and  of excitons
in semiconductors\cite{snoke}.  BEC still persists when a small amount of disorder is added to the system.
BEC in a random environment was observed  in the superfluid  phase of $^4$He in Vycor glass or aerogels\cite{Reppy92}  , in   $^3$He in aerogels\cite{Vincente}
and  in ultra-cold alkali atoms in disordered traps \cite{lye, fallani, schulte, lugan, sanchez, hulet, billy}.
But in a random environment and in the absence of interaction, all Bose-particles fall into the lowest localized single-particle state. Such a ground state is non-ergodic since its
 energy and  spatial extension  depend on a specific realization of the disorder. An arbitrary small repulsive interaction  redistributes the bosons over multiple
potential wells and restores ergodicity.  Hence, contrary to the fermionic case,  the perturbation theory with respect to the interaction strength is
invalid. At low temperature, the Bose   systems display superfluidity provided the density $n$ of  bosons exceeds a critical value $n_c$.
At either  weak disorder or strong interaction, i.e. at $n\gg n_c$, the disorder corrections to
the superfluid density $n_s$ (and the condensate density $n_0$) are small\cite{Huang, vinokur, graham} .
These correction blow up  with the interaction decreasing, signaling the breakdown of the theory.

We present an alternative approach to the
problem of the interaction-induced delocalization starting from deeply localized state of the Bose-gas in a random potential.
We present a simple and visual picture of the deeply localized state, which
decays into remote weakly coupled fragments. We give a geometrical description
of fragments and their distribution in space. At a critical density $n_{c}$, which we express in terms of the
disorder characteristic  and interaction strength, the increasing tunneling of
particles between fragments leads to transition from the random singlet state
to the coherent superfluid.

\section{Single-particle levels in an uncorrelated random
potential\label{single}}

The random environment produces a random potential $U(\textbf{x})$ for the
bosons.  We assume that  $U(\textbf{x})$ is Gaussian distributed with zero average and short range
correlations
\begin{equation}
\left\langle U\left(  \mathbf{x}
\right)  U\left(  \mathbf{x}^{\prime}\right)  \right\rangle =\kappa^2\delta(\textbf{x}-\textbf{x}')
\label{correlator}%
\end{equation}
We will consider briefly the long range correlated case
at the end of this article.
In the absence of interaction the single-particle wave functions obey the    the Schr\"odinger equation
\begin{equation}
\frac{\hbar^{2}}{2m}\nabla^{2}\psi+\left(  E-U\left(  \mathbf{x}\right)
\right)  \psi=0. \label{schroedinger}%
\end{equation}
 Its energy levels
 $E\left[  U\left(  \mathbf{x}\right)  \right]  $
are functionals of the potential $U\left(  \mathbf{x}\right)  $. The
only characteristic of the random potential $\kappa$ together with the Planck's
constant $\hbar$ and the mass $m$ establishes the scales of length and energy:
\begin{equation}\label{larkin}
\mathcal{L}=\frac{\hbar^{4}}{m^{2}\kappa^{2}}, \,\,\,\,\,\,\,\,\,\mathcal{E}=\frac{\hbar^{2}}{m\mathcal{L}^{2}},
\end{equation}
which we call the Larkin length and Larkin energy, respectively \cite{Larkin}. The density of states  $\nu(E)$  belonging to (\ref{schroedinger}) in the limit
$E<0, |E|\gg {\cal E}$ was calculated in \cite{ILifshitz, zittartz, halperin-lax} (for a complete summary see \cite{Lifshitzbook}).
For 3d system
with the volume $\Omega$ it reads:
\begin{equation}
\nu(E)  =\frac{1}{\Omega}\langle\delta\left(  E-E\left[  U\left(
\mathbf{x}\right)  \right]  \right)  \rangle \sim {\cal N}(E)e^{-(|E|/{\cal E})^{1/2}}, \label{DOS}
\end{equation}
where we absorbed a numerical constant in the exponent of $\nu(E)$ in the definition of ${\cal E}$.
As we show below the precise form of the prefactor ${\cal N}(E)$ is not relevant
for our consideration. In a large 3d volume the states with energy $E\gg\mathcal{E}$ are delocalized,
whereas the states with negative energy sufficiently large by modulus $E<0$
and $\left\vert E\right\vert \gg\mathcal{E}$ are strongly localized. The
threshold of localization is a positive energy of the order of $\mathcal{E}$
\cite{remark}. In the interval
between $\mathcal{E}$ and $-\mathcal{E}$ the transition from the extended to
strongly localized states proceeds. The latter are supported by rare
fluctuations of the random potential, which form a potential well sufficiently
deep to have the negative energy $E$ as its only bound state.
Let us introduce the spatial density $n_w(E)$ with the energy less than $E$. It is related to the DOS by equation $n_w(E)=\int_{-\infty}^E \nu (E)dE. $ For deep levels it can be also considered as the spatial density of states
$n_w(R)$ with the radius less than $R$, where $R=\hbar /\sqrt{2m|E|}$. For such states $n_w(R)$ is proportional to a small exponent $\exp(-\sqrt{{\mid
E\mid}/{\cal{E}})}=\exp(-{\cal{L}}/{R})$. From the dimensionality consideration
it follows:
\begin{equation}
n_{w}(E)=R^{-3}f\left(\frac{\cal{L}}{R}\right)\exp\left(-\frac{\cal{L}}{R}\right). \label{n-w}
\end{equation}
The function \(f(x)\)
can be found from Ref.\cite{cardy}
to be proportional to $f(x)\sim x^{\alpha}$ with $\alpha=1$.
It will be inessential for further calculation. The average distance \(d(R)\)
between the wells of the radius less than \(R\) reads: \(d(R)=n^{-1/3}_{w}=Rf^{-1/3}\exp\left(\frac{\cal{L}}{3R}\right)\).
Thus, the distances between the wells are significantly larger than their
sizes. The tunneling factor \(t\left( R \right)\) between two typical wells
with the radius \(R\) of the same order of magnitude is given by a semiclassical
expression \(t\left( R \right)=\exp\left( -\frac{1}{\hbar}\int\left\vert p\right\vert\ dl \right)\), where the path of integration connects the two wells. By the order
of magnitude \(p\sim\hbar/R\) and the length of the integration path is $\sim
d\left( R \right)$. Thus, $\frac{1}{\hbar}\int\left\vert p\right\vert \ dl\approx d/R\approx f^{-1/3} \exp{\left(\frac{\cal{L}}{3R}\right)} $ and
\begin{equation}
t\left( R \right)= \exp\left[ -f^{-1/3}\exp\left(\frac{\cal{L}}{3R}\right) \right].
\label{tunneling-R}
\end{equation}
At $R\sim\mathcal{L}/3$ or $E\sim-9\mathcal{E}$,
the distances between the
optimal potential wells become of the same order of magnitude as their size
$R$. Simultaneously the tunneling amplitude between the wells becomes of the
order of 1. The potential wells percolate and tunneling is not small, but the
states still are not propagating due to the Anderson
localization \cite{Anderson58}.

\section{Bose gas in a large box with an uncorrelated random
potential\label{Gas-in-box}}

In the ground state of an ideal Bose gas in a large box with the Gaussian
random potential all particles are located at the deepest fluctuation level.
In the box of  cubic shape with the side $L$ the deepest level which occurs
with probability of the order of 1 has the radius $R$ determined by equation:
$L^{3}n_{w}\left(  R\right)  =1$, i.e.
$R\sim\frac{\mathcal{L}}{3\ln\left(
L/\mathcal{L}\right)  }$.
The prefactor \(f\) introduces a negligible correction to
the denominator of the order of
$\ln\left(  \ln\frac{L}{\mathcal{L}}\right)$.
The corresponding energy is
$E\sim-9\mathcal{E}\left(
\ln\frac{L}{\mathcal{L}}\right)  ^{2}$.
As we already mentioned such a state is \emph{non-ergodic}
since the location and the depth of the deepest level strongly depends on a
specific realization of the disordered potential. Therefore, the average energy
per particle and other properties averaged over the ensemble has nothing in
common with the properties of a specific sample. Even an infinitely small
repulsion makes the system ergodic in
the thermodynamic limit, i.e. when first the size of the system grows to
infinity and then the interaction goes to zero. In a sufficiently large volume any physical value per particle coincides with its average over the
ensemble. The reason of such a sharp change is that, at any small but finite
interaction, the energy of particles repulsion overcomes their attraction to the potential well when the number of particles increases. They will be
redistributed over multiple wells. Since the distribution of wells in
different parts of sufficiently large volume passes all possible random
configurations with proper ensemble probabilities, the ergodicity is
established. Below we find how the interacting particles eventually
fill localized states.
In a real experiment the Bose gas may be quenched in a metastable state
depending on the cooling rate and other non-thermodynamic factors. This is
what M.P.A. Fisher \textit{et al.} \cite{Fisher} call the Bose glass. Such a
state is also possible in the case of weakly repulsive Bose gas. However, as
it will be demonstrated later, in the case of cooled alkali atoms the
tunneling amplitude still remains large enough to ensure the relaxation to the
equilibrium state in 10$^{-3}\div$10$^{-2}s$. Our further estimates relate to
the real ground state.
As in the Bogolyubov's theory \cite{bogolyubov} 
we assume that the gas criterion $na^{3}\ll1$ is
satisfied. Here $n=N/\Omega$ is the average particle density; $N$ is their
total number and \(a\) is the scattering length. Implicitly our considerations takes in account the change of
the optimal potential well due to the interaction.

Let the Bose gas with the average density of particles $n$ fill all potential
wells with the radii less than $R$ in the ground state. The average number of
particles per well is
%
${N_w}\left(  R\right)  =
{n}/{n_{w}\left(  R\right)  }$.
%
The local density inside the well of the linear size $R$ is $n_{p}%
(R)=\frac{3{ N_w}\left(  R\right)  }{4\pi R^{3}}$. The gain of energy per particle
due to random potential is $E\left(  R\right)  =-\frac{\hbar^{2}}{2mR^{2}}$;
the repulsion energy due to interaction is equal to $gn_{p}\left(  R\right)
=\frac{3\hbar^{2}{N_w}\left(  R\right)  a}{mR^{3}}$, where we used 
the well-known relation for an effective potential field
induced by a gas of scatterers \cite{LLQM}. Minimizing the total energy per
particle $E_{tot}\left(  R\right)  =-\frac{\hbar^{2}}{2mR^{2}}%
+\frac{3\hbar^{2}{N_w}\left(  R\right)  a}{mR^{3}}$ over $R$ 
we find the
value of $R$ corresponding to the minimum of energy at fixed $n$ with the
logarithmic precision:
\begin{equation}
R\left(  n\right)  =\frac{\mathcal{L}}{\ln({n_{c}}/{n})}\,\,\,\,  . \label{R-n}%
\end{equation}
%
$\textrm{where}\,\,\,\,n_{c}=\left(  3\mathcal{L}^{2}a\right)  ^{-1}$ denotes the critical density. 
The factor $f$ in equation (\ref{n-w}) leads to  corrections of the type
$\ln\left(  \ln({n_{c}}/{n})\right)  $
 which can be neglected. Further we put $f=1$. The distances
between the filled wells according to the corresponding expression \(d(R)\)
for single-particle states reads  
$d\left(  n\right)  ={\mathcal{L}}({\ln({n_{c}}/{n})})^{-1}\left(  {n_{c}/%
}{n}\right)  ^{1/3}$. 
They strongly exceed the average size of the potential well
(\ref{R-n}) at $n\ll n_{c}$. At the same condition the chemical potential of
atoms can be estimated as 
$\mu\left(  n\right)  =-\frac{\hbar^{2}}{2mR^{2}\left(  n\right)  }%
=-\frac{\mathcal{E}}{2}\left(  \ln\frac{n_{c}}{n}\right)  ^{2}$.
The tunneling amplitude $t\left(  n\right)  $ between two wells separated by a
typical distance $d\left(  n\right)  $ can be found by employing the single
particle result (\ref{tunneling-R}):
\begin{equation}
t\left(  n\right)  =\exp\left[  -\left(  {n_{c}}/{n}\right)
^{1/3}\right]  . \label{tunneling-n}%
\end{equation}
Thus, the Bose gas at $n\ll n_{c}$ is fragmented into multiple clusters of
small size $R\left(  n\right)  $ separated by much larger distances
$d\left(
n\right)  $ and containing about $\mathcal{L}/\left[  3a\left(  \ln\frac
{n_{c}}{n}\right)  ^{3}\right]  $
 particles each.
 The amplitude of tunneling
between the wells depends on the scattering length  in a non-analytic way
and is
exponentially small for weak interaction. Therefore, the number of particles in
each cluster is well defined. As a consequence, the phase is completely uncertain.
Such a state is a singlet with non-uniformly distributed particles, a random
singlet: the ground state is non-degenerate.
 %
 The compressibility $\frac{\partial n}{\partial \mu}=
 \frac{n}{\cal E}\ln\frac{n_c}{n}$ is finite as expected for the Bose glass phase \cite{Fisher}.

\section{Bosons in atomic traps}

Our results can be easily extended to bosons in harmonic traps characterized by a potential
\begin{equation}
V_{trap}=\frac{m\omega^{2}R^{2}}{2}=\frac{\hbar^{2}}{2m}\frac{R^{2}}{\ell^{4}} \label{trap-R}%
\end{equation}
where we introduced the oscillator length $\ell=\sqrt{\hbar/(m\omega)}$.
This section partly overlaps sligthly with our previous work \cite{PRL}.
The energy of the bosons includes now four terms: the kinetic energy, the
confining potential energy of the trap, the repulsion from other particles and
the energy of the random potential. Two of them, the interaction with the trap
and the random potential tend to confine and localize the particle.
Going through essentially the same steps as before, we can distinguish four different regimes (Figure 1).
\begin{figure}
\begin{center}
\includegraphics[width=8.5cm,height=6.7cm]{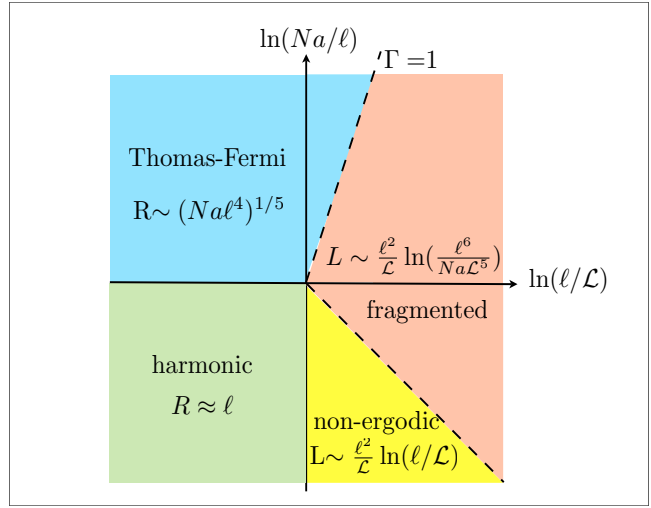}
\caption{{Regime diagram of atoms in traps: uncorrelated disorder.  $R$ denotes the size of the single existing atomic cloud. $L$ is the size of the cloud of fragments.}}
\label{Fig:regime_diagram_uncorrelated}
\end{center}
\end{figure}

\noindent1. \emph{Weak disorder and weak interaction}: $3Na\ll \ell \ll\mathcal{L}$ .
In this case the
interaction can be neglected. Minimizing the remaining terms, the kinetic
energy and energy of the trap, we find $R=\ell$. Physically it means that all
particles are condensed in the oscillator ground state.

\noindent2. \textit{Weak disorder and strong  interaction: } $\ell \ll\mathcal{L} ,\,\, \ell\ll3Na$ .
Neglecting the kinetic
energy and minimizing remaining energy of traps plus the repulsion energy, one
finds the result known as Thomas-Fermi approximation\cite{Dalfovo}: $R=\left(
\frac{9}{2}Na\ell^{4}\right)  ^{1/5}$.


\noindent3. \emph{Strong disorder and weak interaction :}  $3Na\ll\mathcal{L}\ll\ell$ .
In this
range of variables the non-ergodic phase is realized. Since interaction is
negligible, the particles find a random potential well with the deepest level
and fall into it. Let such a well can be found at a distance $\sim L$ from the
trap center. Its depth typically is about $9\mathcal{E}\ln^{2}\left(
L/\mathcal{L}\right)  $. This gain of energy must be not less than the loss of
the trap energy $m\omega^{2}L^{2}/2$. A typical value of $L$ appears when both
this energies have the same order of magnitude. Thus, $L\approx6\sqrt{2}\left(
\ell^{2}/\mathcal{L}\right)  \ln\left(  \ell/\mathcal{L}\right)  $. A typical
size of the well is $R\approx\mathcal{L}/\left(  6\ln\left(  \ell
/\mathcal{L}\right)  \right)  $.

\noindent4. \emph{Strong disorder and moderate interaction: } $ \mathcal{L}\ll  3Na\ll\ell$. In
this case the ergodicity is restored. Our experience with the gas in a box
prompts that the gas cloud is split into fragments each occupying a random
potential well from very small size till same size $R$ depending on $N$.
The typical disorder energy per particle is
$\mu=-\mathcal{E}\left(  \ln\frac{n_{c}}{n}\right)  ^{2}$. It becomes equal to
the trap energy at the distance $L\sim\left(  \ell^{2}/\mathcal{L}\right)
{\ln\Gamma}$
where $\Gamma$ is a new dimensionless parameter
\begin{equation}\Gamma=\frac{\ell^{6}}{3Na\mathcal{L}^{5}}\sim \frac{ n_c}{n}.
\end{equation}\label{Gamma}
Therefore, the average density is $n\sim N\mathcal{L}^{3}/\ell^{6}$. The state of the Bose gas is fragmented and strongly
localized when $\Gamma$ is large; the transition to delocalized superfluid state
proceeds when this ratio becomes \(\sim\)1.
The phase diagram is shown in Fig. \ref{Fig:regime_diagram_uncorrelated}.
Note the
counter-intuitive dependence of the size on the number of particles: the cloud
slightly contracts with increasing number of particles. It happens because the
number of particles in each fragment increases more rapidly with the average
density than the number of fragments.

\section{Correlated disorder}

So far we considered uncorrelated disorder (\ref{correlator}). Our results can be extended to
random potentials with a finite correlation length $b$ and strength %
$U_0=\sqrt{\langle U^{2}\left( \bf{x} \right)\rangle}$.
We quote here the results without derivation, which can be found in Ref. \cite{long_paper}.
As long as $b\ll {\cal L}=\frac{3\hbar^4}{4\pi m^2U_{0}^2b^{3}}$ the results of the previous considerations remain correct.
 In the opposite case the optimal potential wells have the width \(b\) and, contrary to the short range
 correlated case,  they contain many bound states (of the order of $(b/{\cal L})^{3/4}\gg 1$).
 It is convenient to introduce a new length scale
\(B=b\left({\cal L}/{b}\right)^{1/4}=\left({3}/{4\pi}\right)^{1/4}\left({\hbar^2}/
{mU_0}\right)^{1/2}\). In the following we restrict our consideration to the case $b\gg {\cal L}$, i.e. $b\gg B$. The density of states
in this case is $\nu(E) \sim \exp [{-(E/U_0)^2}]$ \cite{Lifshitzbook}. 
The critical density is  $n_c\sim 1/(aB^2)$ 
(this result has been found before in \cite{Shlo})
and the typical size of a fragment is \( R\approx b (\ln(n_c/n))^{-1/2}\). The  distance between fragments is $d(n)\approx b(n_c/n)^{1/3}$
and the tunneling coefficient is $t(n)\approx \exp[-(b/B)(n_c/n)^{1/3}]$.

In the case of a harmonic trap we again find four different
regimes (Fig. 2). The relevant parameter is $\Gamma=\ell^6/(NaB^5)\approx n_c/n$.
For $\Gamma\approx 1$ the transition to the superfluid phase proceeds.
\begin{figure}
\begin{center}
\includegraphics[width=8.5cm,height=6.7cm]{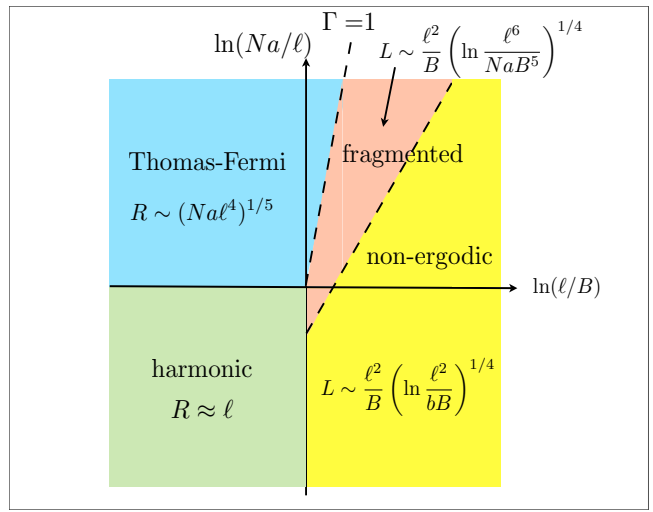}
\caption{\small{Regime diagram of atoms in traps: correlated disorder.  $R$ denotes the size of the single atomic cloud, $L$ denotes the size of the fragmented state.}}
\label{default}
\end{center}
\end{figure}
All results can be extended to lower dimensions \cite{long_paper}.
\section{Conclusions}

Four parameters can be controllably and independently varied in the
experiment. They are: number of particles $N$; the frequency $\omega$ or
equivalently the strength of the trap; the scattering length $a$ (it can
be varied by
approaching one of the Feshbach resonances); the strength of disorder $U_{0}$.
Using this freedom it is feasible to pass all regimes described
above.
A simple estimate shows that, at
$b\sim1\mu m$, the transition from uncorrelated to strongly correlated regime
proceeds at frequency of disorder potential $\omega_{d}=\sqrt
{2U_{0}/mb^{2}}\sim1kHz$ which is accessible.

Simplest experiments are the measurements of the cloud size $L$ as a function
of different variable parameters in the regime of multiple localized
fragments.
Theory predicts that in the regime of uncorrelated disorder the size of the
cloud is proportional to \(U^{2}_{0}/\omega\) . It also predicts very
weak dependence of the size on the number of particles $\sim{\ln N}$. In
the case of strongly correlated disorder the size of the cloud is proportional to $\omega U_{0}^{1/2}$; the dependence on $N$ also is weaker
than in the uncorrelated regime: $L\propto\left(  \ln N\right)  ^{1/4}$. \

It would be important to observe the transition from non-ergodic state with
one or few fragments to the ergodic state with many fragments and check that
it happens at $N=\mathcal{L}/3a$ for uncorrelated disorder and at
$N=\left({b^3}/{3aB^2}\right) $
for strongly correlated disorder.

Another feasible experiment is the time-of-flight spectroscopy after switching
off both the trap and the random potential. In this experiment the
distribution of particles over momenta (velocities) is measured. Its width $\Delta p$ is associated with the
average size of the fragment $R $ by the uncertainty relation \(\Delta p=\hbar/R\).
It gives the opportunity to check the equation \(R=\mathcal{L}/\ln\Gamma\) for the uncorrelated
disorder or $R=\mathcal{L}/\ln\Gamma$ for correlated disorder. Installing a counter close to the trap, at a distance comparable to the size of
the trap, would allow to register the oscillations of the particle flux due to
discrete character of the fragmented state. This is an opportunity to
find the distances between fragments and compare theory with experiment.

The transition between localized and delocalized coherent state in the random
potential was found in several experiments (see Introduction). We propose
to make more detailed measurement of the transition manifold and check our
predictions.

An important question is whether the relaxation to the ground state can be
reached during a reasonable time interval compatible with the time of
experiment. We analyze this question for the uncorrelated or weakly correlated
disorder. In this case the relaxation time due to tunneling can be estimated
as $\tau=2\pi\omega_{n}^{-1}t^{-1}$, where $\omega_{n}\sim\frac{\mathcal{E}
}{\hbar}\left(  \ln\Gamma\right)  ^{2}$ is the characteristic frequency of
the optimal potential well and $t\sim\exp\left[  -\Gamma^{1/3}\right]  $ is
the tunneling coefficient (see eq. (\ref{tunneling-n})). For numerical
estimates we accept $\Gamma\sim125$, $\ell\sim10\mu m$, $b\sim\mathcal{L}
\sim1\mu m$, $a\sim0.01\mu m$, $N\simeq27,000$. Then $t^{-1}=148$ and
$\tau\sim0.06s$. The Larkin length can be increased by decreasing the
amplitude of the random potential.
Simultaneously, at fixed values $N$, $\ell$ and $a,$ the value $\Gamma$
decreases as $\mathcal{L}^{-5}$. This example shows that the equilibrium is
accessible, though it is difficult to reach large ratio $\mathcal{L}/b$.

The closest to ours was the approach developed in the work Lugan et al. \cite{lugan}. Apart
from the fact that these authors considered only the 1-dimensional case, the main
difference between our and their problems is that they considered the random 
potential with the exact lower boundary $U_{\rm b}$ and with on-site distribution function  
$ W[U]\propto\exp[-c(U-U_{\rm b})]$   instead of a Gaussian distribution.  Such a distribution 
allows deeply localized states  only at energies $E$ close to the exact lower boundary $U_{\rm b}$. 
The corresponding fluctuations have the width R the broader the closer is $E$ to $U_{\rm b}$. 
It is clear that these levels are very different from those discussed above. If the random 
potential has the exact lower boundary, our theory is valid only if this boundary 
is separated from the most probable value of the potential by an energy interval strongly 
exceeding the energy dispersion. Then the localized states of our theory appear at 
intermediate energies between dispersion and $U_{\rm b}$.


\acknowledgments
The authors acknowledge a helpful support from the DFG through NA222/5-2 and
SFB 680/D4(TN) as well as from the DOE under the grant DE-FG02-06ER46278.


\begin{thebibliography}{0}

\bibitem{Anderson58}
  \Name{Anderson P. W.}
  \REVIEW{Phys. Rev.}{109}{1958}{1492}.

\bibitem{gangof4}
  \Name{Abrahams E., Anderson P. W., Licciardello D. C. \and Ramakrishnan T. V.}
  \REVIEW{Phys. Rev. Lett.}{42}{1978}{673}.

\bibitem{review}
  \Name{Lee A. \and Ramakrishnan T. V.}
  \REVIEW{Rev. Mod. Phys.}{57}{1985}{287}.

\bibitem{ILifshitz}
  \Name{Lifshitz I. M.}
  \REVIEW{Sov. Phys. JETP}{26}{1968}{462}.

\bibitem{zittartz}
  \Name{Zittartz J. \and Langer J. }
  \REVIEW{Phys. Rev.}{148}{1966}{741}.

\bibitem{halperin-lax}
  \Name{Halperin B. I. \and Lax M.}
  \REVIEW{Phys. Rev.}{153}{1966}{802}.

\bibitem{altshuler+aronov}
  \Name{Altshuler B. L. \and Aronov A. G.}
  \REVIEW{Sov. Phys. JETP}{50}{1979}{968}.

\bibitem{finkelstein}
  \Name{Finkelstein A. M.}
  \REVIEW{Sov. Phys. JETP}{57}{1983}{97}.

\bibitem{kravchenko-b}
  \Name{Abrahams E., Kravchenko S. V. \and Sarachik M. P.}
  \REVIEW{Rev. Mod. Phys.}{73}{2001}{251}.







\bibitem{Leggett}
  \Name{Leggett A. J.}
  \REVIEW{Rev. Mod. Phys.}{73}{2001}{307}.

\bibitem{Thinkham}
  \Name{Schrieffer J. R. \and Tinkham M.}
  \REVIEW{Rev. Mod. Phys.}{71}{1999}{313}.


\bibitem{Ketterle}
  \Name{Ketterle W.}
  \REVIEW{Rev. Mod. Phys.}{74}{2000}{1131}.

\bibitem{Dalfovo}
  \Name{Dalfovo F. D., Giorgini S., Pitaevskii L. P. \and Stringari S.}
  \REVIEW{Rev. Mod. Phys.}{71}{1999}{463}.

\bibitem{snoke}
  \Name{Snoke D.}
  \REVIEW{Science}{298}{2002}{1368}.

\bibitem{Reppy92}
  \Name{Reppy J. D.}
  \REVIEW{J. Low T. Phys.}{87}{1992}{205}.

\bibitem{Vincente}
  \Name{Vicente C. L., Choi H. C., Xia J. S., Halperin W. P., Mulders N. \and Lee Y.}
  \REVIEW{Phys. Rev. B}{72}{2005}{094519}.

\bibitem{lye}
  \Name{Lye J. E., Fallani L., Modugno M., Wiersma D. S., Fort C. \and Inguscio M.}
  \REVIEW{Phys. Rev. Lett.}{95}{2005}{070401}.

\bibitem{schulte}
  \Name{Schulte T., Drenkelforth S., Kruse J., Ertmer W., Arlt J., Sacha K.,
  Zakrzewski J. \and Lewenstein M.}
  \REVIEW{Phys. Rev. Lett.}{95}{2005}{170411}.

\bibitem{fallani}
  \Name{Fallani L., Lye J. E., Guarrera V., Fort C., \and Inguscio M.}
  \REVIEW{Phys. Rev. Lett.}{98}{2007}{130404}.

\bibitem{lugan}
  \Name{Lugan P., Clement D., Bouyer P., Aspect A., Lewenstein M. \and Sanchez-Palencia L.}
  \REVIEW{Phys. Rev. Lett.}{98}{2007}{170403}.

\bibitem{sanchez}
  \Name{Sanchez-Palencia L., Clement D., Lugan P., Bouyer P., Shlyapnikov G. V. \and Aspect A.}
  \REVIEW{Phys. Rev. Lett.}{98}{2007}{210401}.

\bibitem{hulet}
  \Name{Chen Y. P., Hitchcock J., Dries D., Junker M., Welford C. \and Hulet R. G.}
  \REVIEW{Phys. Rev. A}{77}{2008}{033632}.

\bibitem{billy}
  \Name{Billy J., Josse V., Zuo Z., Bernard A., Hambrecht B., Lugan P., Clement D., Sanchez-Palencia L., Bouyer P. \and Aspect A.}
  \REVIEW{Nature}{453}{2008}{891}.

\bibitem{Huang}
  \Name{Huang K., Meng H. F.}
  \REVIEW{Phys. Rev. Lett.}{69}{1992}{644}.

\bibitem{vinokur}
  \Name{Lopatin A. V. \and V. M. Vinokur}
  \REVIEW{Phys. Rev. Lett.}{88}{2002}{235503}.

\bibitem{graham}
  \Name{Falco G. M., Pelster A. \and Graham R.}
  \REVIEW{Phys. Rev. A}{75}{2007}{063619}.

\bibitem{Larkin}
  \Name{Larkin A. I.}
  \REVIEW{Sov. Phys. JETP}{31}{1970}{784}.

\bibitem{Lifshitzbook}
  \Name{Lifshitz I. M., Gredeskul S. A. \and Pastur L. A.}
  \Book{Introduction to the theory of disordered systems}
  \Publ{Wiley-Interscience, New York}
  \Year{1988}.

\bibitem{remark} This is correct only for 3d systems. As it was conjectured in the
work by Abrahams et al. \cite{gangof4}, all single-particle states in 1 and
2d systems are localized.

\bibitem{cardy}
  \Name{Cardy J.}
  \REVIEW{J. Phys. C: Solid State Phys.}{11}{1978}{L321}.

\bibitem{Fisher}
  \Name{Fisher M. P. A., Weichman P. B., Grinstein G. \and Fisher D. S.}
  \REVIEW{Phys. Rev. B}{40}{1989}{546}.

\bibitem{bogolyubov}
  \Name{Bogoliubov N. N.}
  \REVIEW{J. Phys. USSR}{11}{1947}{23}.

\bibitem{LLQM}
  \Name{Landau L. D. \and Lifshitz E. M.}
  \Book{Quantum Mechanics, $3^{rd}$ ed.}
  \Publ{Pergamon}
  \Year{1991}.

\bibitem{PRL}
  \Name{Nattermann T. \and Pokrovsky V. L.}
  \REVIEW{Phys. Rev. Lett.}{100}{2008}{060402}.

\bibitem{Shlo}
  \Name{Shklovskii B. I.}
  \REVIEW{Semiconductors (St. Peterburg)}{42}{2008}{927}.


\bibitem{long_paper}
  \Name{Falco G. M., Nattermann T. \and Pokrovsky V. L.}
  \REVIEW{arXiv:0811.1269}{}{2008}{}.

\end{thebibliography}
\end{document}